\documentclass[sigconf]{acmart}

\usepackage{graphicx}
\usepackage{algorithm} 
\usepackage{algpseudocode}
\usepackage{booktabs} 
\usepackage{amsthm} 
\usepackage{bbm} 
\usepackage{tabularx} 
\usepackage[flushleft]{threeparttable} %

\usepackage[flushleft]{threeparttable} 
\usepackage{booktabs,caption}

\theoremstyle{remark}

\begin{document}

\copyrightyear{2024}
\acmYear{2024}
\setcopyright{acmlicensed}\acmConference[IDE '24]{2024 First IDE Workshop}{April 20, 2024}{Lisbon, Portugal}
\acmBooktitle{2024 First IDE Workshop (IDE '24), April 20, 2024, Lisbon, Portugal}
\acmDOI{10.1145/3643796.3648455} \acmISBN{979-8-4007-0580-9/24/04}

\title {JetTrain: IDE-Native Machine Learning Experiments}

\author{Artem Trofimov}
\affiliation{%
  \institution{JetBrains}
  \city{Berlin}
  \country{Germany}}
\email{artem.trofimov@jetbrains.com}

\author{Mikhail Kostyukov}
\affiliation{%
  \institution{JetBrains}
  \city{Amsterdam}
  \country{Netherlands}}
\email{mikhail.kostyukov@jetbrains.com}

\author{Sergei Ugdyzhekov}
\affiliation{%
  \institution{JetBrains}
  \city{Munich}
  \country{Germany}}
\email{sergei.ugdyzhekov@jetbrains.com}

\author{Natalia Ponomareva}
\affiliation{%
  \institution{JetBrains}
  \city{Berlin}
  \country{Germany}}
\email{natalia.ponomareva@jetbrains.com}

\author{Igor Naumov}
\affiliation{%
  \institution{JetBrains}
  \city{Belgrade}
  \country{Serbia}}
\email{igor.naumov@jetbrains.com}

\author{Maksim Melekhovets}
\affiliation{%
  \institution{JetBrains}
  \city{Berlin}
  \country{Germany}}
\email{maksim.melekhovets@jetbrains.com}

\begin{abstract}

Integrated development environments (IDEs) are prevalent code-writing and debugging tools. However, they have yet to be widely adopted for launching machine learning (ML) experiments. This work aims to fill this gap by introducing JetTrain, an IDE-integrated tool that delegates specific tasks from an IDE to remote computational resources. A user can write and debug code locally and then seamlessly run it remotely using on-demand hardware. We argue that this approach can lower the entry barrier for ML training problems and increase experiment throughput. 

\end{abstract}




\maketitle

\section {Introduction}
\label{jettrain-introduction}

One of the core parts of a machine learning workflow is training. Training is adjusting model internal parameters (like weights in a neural network) to minimize errors in predictions or decisions. Multiple training runs form an {\em experiment} that checks some hypothesis about a model improvement. 

Training or fine-tuning modern machine learning models requires complex hardware, especially in the LLM era~\cite{isaev2023scaling}. Thus, ML engineers use various computational resources for code writing and experiment launching. This leads to overcomplicated ML experimentation tools requiring context switching~\cite{quaranta2021taxonomy}.

This work proposes a novel IDE-integrated approach to launching ML experiments called {\em JetTrain}. We hypothesize that it can lower the entry barrier for users familiar with IDE and decrease the adverse effects of context switching. We overview existing interfaces for launching ML experiments in Section~\ref{jettrain-motivation}. Our approach is introduced in Section~\ref{jettrain-experiments}, and the challenges are discussed in Section~\ref{jettrain-challenges}.

\section {Motivation}
\label{jettrain-motivation}

There are multiple interfaces to launch ML experiments on remote hardware. In this section, we discuss widely adopted approaches and highlight their advantages and limitations.

\begin{table*}[t!]
\begin{threeparttable}
  \begin{tabularx}{\linewidth}{@{\extracolsep{\fill}}lccccc}
  Tool/Approach & Cost efficiency & Reproducibility & Onboarding efficiency & Context persistence & Debug capabilities \\
  \hline
  SSH                 & low  & low  & high & high$^*$ & high \\
  Jupyter Notebooks   & low  & low  & high & high & medium \\
  Pipeline Tools      & high & high & low  & low & low \\
  Task Schedulers     & high & high & medium & medium & low \\
  JetTrain            & high & high & high & high & high \\
  \hline
  \end{tabularx}
  * with remote development
  \end{threeparttable}
  \caption{Interfaces for launching ML experiments}
  \label{table-competitors}
\end{table*}

{\bf Secure Shell (SSH)} connection to rented virtual machines (VMs) or on-premise servers is the most straightforward approach to launch experiments. An ML engineer should install all required libraries, download data, and run a locally prepared code. Using remote development features in an IDE for these purposes is even possible. Nonetheless, this approach exhibits limited scalability and is ineffective in cost.

{\bf Jupyter Notebooks} provide complete control over code execution; they are highly customizable and integrated with IDE. Notebooks are great for analysts and data scientists due to their rich visualization features. Nevertheless, code in notebooks is usually poorly reproducible, testable, and hard to deploy~\cite{pimentel2021understanding}. Another problem is that notebooks are not designed to ensure efficient hardware utilization and are ineffective in cost.

{\bf Pipeline Tools}, including but not limited to KF Pipelines~\cite{bisong2019kubeflow}, Ray~\cite{moritz2018ray}, Metaflow~\cite{tagliabue2023reasonable}, and others, primarily facilitate the productionalization of processes but lacks debug functionality. These tools require ML engineers to adjust their code to specific frameworks, raising the entry barrier. Not every experiment reaches production; these tools bring production complexities into the experimentation stage, possibly reducing the number of experiments.

{\bf Task Scheduling Tools}, such as SkyPilot~\cite{yang2023skypilot}, DStack~\cite{dstack}, MosaicML~\cite{mosaicml}, and others, are the last but not least group in our overview. ML engineers do not need to change an existing code; they can write a YAML configuration and run an experiment with a CLI interface. YAML with CLI interface is questionable for the experimentation scenario because it lacks remote debugging, terminal, and other IDE-provided features.

Each tool in the list has specific advantages and limitations summarized in Table~\ref{table-competitors}. As we can see, there is a gap between simple interfaces (SSH and Jupyter) and more sophisticated alternatives (Pipeline and Task Scheduling Tools). The first group ensures high onboarding efficiency and context persistence, while the second group provides efficient hardware utilization and reproducibility.

We see a promising direction in filling this gap with a tool integrated into IDE. We aim to provide simple UX without the need for context switching on the one side and a mature scheduler tool under the hood on the other side. The following section details this idea.

\section {IDE-native ML experiments}
\label{jettrain-experiments}

We aim to provide a plugin and an underlying service that ensures a smooth transition between IDE and remote experiments. 
All development remains local, and computational resources are allocated only for experiments on demand. 
Therefore, a user does the following actions within launching an ML experiment scenario:

\begin{enumerate}
  \item Opens an IDE
  \item Runs and debugs code locally
  \item Writes command to launch an experiment
  \item Chooses hardware setup capable of running an experiment
  \item Indicates data to mount if needed remotely
  \item Launches the experiment on a remote hardware
  \item Starts debug session (optional)
  \item Connects to a remote executor with terminal (optional)
\end{enumerate}

To create such a tool, we need to extend an IDE with a particular type of execution. Fortunately, even in a local IDE run configuration, we have almost all the required information to migrate the run to a remote machine: a working directory to synchronize, an environment to install, and a command to run. We must add only a few additional properties, such as external data to mount and hardware provisioning parameters (GPU type, number of GPUs, etc.).

This approach has a low entry barrier and preserves the working context because a user does not need to leave an IDE. Other benefits we obtain are development infrastructure features already integrated in IDEs such as debugger, terminal, etc.

We need an integration with some cloud scheduler to ensure efficient hardware utilization. In the scheduler, we need support for custom protocol connections to maintain the debug and terminal protocols. 
Another requirement is mounting data from user-defined storages, e.g., S3, so the user can bring its data for training. 
We use TeamCity~\cite{mahalingam2014learning} as a cloud scheduler because it has various cloud connectors and is optimized for long-running tasks. 

Several underlying implementation challenges need to be clarified: efficiently synchronizing user code and data from a local machine, ensuring reproducibility, and supporting asynchronous debugging. We cover these topics in the following section.

\section {Challenges}
\label{jettrain-challenges}

In this section, we discuss the main technical challenges that we face during JetTrain implementation.

{\bf Code and data sychronization} is required to run an experiment using the same code as on a local machine. There are multiple ways to do that. Skypilot CLI client solves this task using {\em rsync} via SSH connection~\cite{yang2023skypilot}. This solution works well if local and remote machines are geographically close. However, in real-life scenarios, a remote machine can be far away from a local one so that the internet connection can limit the synchronization throughput. To overcome this issue, we propose to use geographically distributed storage like Amazon S3 for data synchronization. Firstly, data is synchronized to the closest bucket to a local machine and then transferred to the target geographical zone using a CDN service.

{\bf Reproducibility} is one of the essential requirements for ML experiments. ML engineers usually persist experiment metrics using special tracking tools like Weights and Biases~\cite{wandb} or MLFlow~\cite{zaharia2018accelerating}, but there is also a need to restore the experiment's code. One way to do that is to attach a git commit revision to an experiment, but it can be inconvenient for parameter selection tasks. It also may lead to an enormous number of meaningless commits. As stated earlier, we should synchronize user code and data from a local machine. We reuse this behavior for reproducibility as well: each run forms the snapshot of a working directory. To minimize network traffic from a local machine, we can copy data from a previous snapshot first and then upload local changes. 

{\bf Asynchronous debug} is required when a remote machine can be allocated sometime after a user starts the debug session. It may occur due to the GPU shortage in cloud providers or limited on-premises resources. Hence, a user can set a breakpoint and click the debug button, but the actual debug session may start several hours after the task is scheduled. To support this scenario, we must buffer debug protocol messages and reconnect to the existing remote debug session. We do not have a satisfactory solution for this problem; the solution may even require changes in the debug protocol.  

\section{Conclusion and Future Work}
\label{jettrain-conclusion}

In this work, we introduced JetTrain - a tool that integrates machine learning experimentation with coding environments, aiming to make ML training more accessible and efficient for developers. The main idea behind JetTrain is to keep the development process local but delegate runs of experiments to a variety of remote computational resources on-demand, ensuring local-like UX (debugger, terminal, etc.). This tool can make ML models training UX significantly more straightforward than state-of-the-art competitors.

We demonstrated the challenges and pitfalls one may face while implementing the JetTrain concept. Data and environment synchronization, experiment reproducibility, and asynchronous debugging are among them. We highlighted techniques that may help us overcome these issues. In our future work, we plan to share the performance metrics of our solution and show how it affects the productivity of ML teams.

\bibliographystyle{ACM-Reference-Format}
\bibliography{jettrain}


\begin{thebibliography}{12}


\ifx \showCODEN    \undefined \def \showCODEN     #1{\unskip}     \fi
\ifx \showDOI      \undefined \def \showDOI       #1{#1}\fi
\ifx \showISBNx    \undefined \def \showISBNx     #1{\unskip}     \fi
\ifx \showISBNxiii \undefined \def \showISBNxiii  #1{\unskip}     \fi
\ifx \showISSN     \undefined \def \showISSN      #1{\unskip}     \fi
\ifx \showLCCN     \undefined \def \showLCCN      #1{\unskip}     \fi
\ifx \shownote     \undefined \def \shownote      #1{#1}          \fi
\ifx \showarticletitle \undefined \def \showarticletitle #1{#1}   \fi
\ifx \showURL      \undefined \def \showURL       {\relax}        \fi
\providecommand\bibfield[2]{#2}
\providecommand\bibinfo[2]{#2}
\providecommand\natexlab[1]{#1}
\providecommand\showeprint[2][]{arXiv:#2}

\bibitem[\protect\citeauthoryear{Biewald}{Biewald}{2023}]%
        {wandb}
\bibfield{author}{\bibinfo{person}{Lukas Biewald}.} \bibinfo{year}{2023}\natexlab{}.
\newblock \bibinfo{title}{Experiment Tracking with Weights and Biases}.
\newblock   (\bibinfo{year}{2023}).
\newblock
\showURL{%
\url{https://www.wandb.com/}}
\newblock
\shownote{Software available from wandb.com.}


\bibitem[\protect\citeauthoryear{Bisong and Bisong}{Bisong and Bisong}{2019}]%
        {bisong2019kubeflow}
\bibfield{author}{\bibinfo{person}{Ekaba Bisong} {and} \bibinfo{person}{Ekaba Bisong}.} \bibinfo{year}{2019}\natexlab{}.
\newblock \showarticletitle{Kubeflow and kubeflow pipelines}.
\newblock \bibinfo{journal}{{\em Building Machine Learning and Deep Learning Models on Google Cloud Platform: A Comprehensive Guide for Beginners\/}} (\bibinfo{year}{2019}), \bibinfo{pages}{671--685}.
\newblock


\bibitem[\protect\citeauthoryear{dstack}{dstack}{2023}]%
        {dstack}
\bibfield{author}{\bibinfo{person}{dstack}.} \bibinfo{year}{2023}\natexlab{}.
\newblock \bibinfo{title}{An easier way to train and deploy generative AI}.
\newblock \bibinfo{howpublished}{\url{https://dstack.ai}}.   (\bibinfo{year}{2023}).
\newblock


\bibitem[\protect\citeauthoryear{Isaev, McDonald, and Vuduc}{Isaev et~al\mbox{.}}{2023}]%
        {isaev2023scaling}
\bibfield{author}{\bibinfo{person}{Mikhail Isaev}, \bibinfo{person}{Nic McDonald}, {and} \bibinfo{person}{Richard Vuduc}.} \bibinfo{year}{2023}\natexlab{}.
\newblock \showarticletitle{Scaling Infrastructure to Support Multi-Trillion Parameter LLM Training}. In \bibinfo{booktitle}{{\em Architecture and System Support for Transformer Models (ASSYST@ ISCA 2023)}}.
\newblock


\bibitem[\protect\citeauthoryear{Mahalingam}{Mahalingam}{2014}]%
        {mahalingam2014learning}
\bibfield{author}{\bibinfo{person}{Manoj Mahalingam}.} \bibinfo{year}{2014}\natexlab{}.
\newblock \bibinfo{booktitle}{{\em Learning Continuous Integration with TeamCity}}.
\newblock \bibinfo{publisher}{Packt Publishing Ltd}.
\newblock


\bibitem[\protect\citeauthoryear{Moritz, Nishihara, Wang, Tumanov, Liaw, Liang, Elibol, Yang, Paul, Jordan, et~al\mbox{.}}{Moritz et~al\mbox{.}}{2018}]%
        {moritz2018ray}
\bibfield{author}{\bibinfo{person}{Philipp Moritz}, \bibinfo{person}{Robert Nishihara}, \bibinfo{person}{Stephanie Wang}, \bibinfo{person}{Alexey Tumanov}, \bibinfo{person}{Richard Liaw}, \bibinfo{person}{Eric Liang}, \bibinfo{person}{Melih Elibol}, \bibinfo{person}{Zongheng Yang}, \bibinfo{person}{William Paul}, \bibinfo{person}{Michael~I Jordan}, {et~al\mbox{.}}} \bibinfo{year}{2018}\natexlab{}.
\newblock \showarticletitle{Ray: A distributed framework for emerging $\{$AI$\}$ applications}. In \bibinfo{booktitle}{{\em 13th USENIX symposium on operating systems design and implementation (OSDI 18)}}. \bibinfo{pages}{561--577}.
\newblock


\bibitem[\protect\citeauthoryear{Pimentel, Murta, Braganholo, and Freire}{Pimentel et~al\mbox{.}}{2021}]%
        {pimentel2021understanding}
\bibfield{author}{\bibinfo{person}{Jo{\~a}o~Felipe Pimentel}, \bibinfo{person}{Leonardo Murta}, \bibinfo{person}{Vanessa Braganholo}, {and} \bibinfo{person}{Juliana Freire}.} \bibinfo{year}{2021}\natexlab{}.
\newblock \showarticletitle{Understanding and improving the quality and reproducibility of Jupyter notebooks}.
\newblock \bibinfo{journal}{{\em Empirical Software Engineering\/}} \bibinfo{volume}{26}, \bibinfo{number}{4} (\bibinfo{year}{2021}), \bibinfo{pages}{65}.
\newblock


\bibitem[\protect\citeauthoryear{Quaranta, Calefato, and Lanubile}{Quaranta et~al\mbox{.}}{2021}]%
        {quaranta2021taxonomy}
\bibfield{author}{\bibinfo{person}{Luigi Quaranta}, \bibinfo{person}{Fabio Calefato}, {and} \bibinfo{person}{Filippo Lanubile}.} \bibinfo{year}{2021}\natexlab{}.
\newblock \showarticletitle{A taxonomy of tools for reproducible machine learning experiments}.
\newblock \bibinfo{journal}{{\em AIxIA 2021\/}} (\bibinfo{year}{2021}).
\newblock


\bibitem[\protect\citeauthoryear{Tagliabue, Bowne-Anderson, Tuulos, Goyal, Cledat, and Berg}{Tagliabue et~al\mbox{.}}{2023}]%
        {tagliabue2023reasonable}
\bibfield{author}{\bibinfo{person}{Jacopo Tagliabue}, \bibinfo{person}{Hugo Bowne-Anderson}, \bibinfo{person}{Ville Tuulos}, \bibinfo{person}{Savin Goyal}, \bibinfo{person}{Romain Cledat}, {and} \bibinfo{person}{David Berg}.} \bibinfo{year}{2023}\natexlab{}.
\newblock \showarticletitle{Reasonable Scale Machine Learning with Open-Source Metaflow}.
\newblock \bibinfo{journal}{{\em arXiv preprint arXiv:2303.11761\/}} (\bibinfo{year}{2023}).
\newblock


\bibitem[\protect\citeauthoryear{Team}{Team}{2023}]%
        {mosaicml}
\bibfield{author}{\bibinfo{person}{The Mosaic~ML Team}.} \bibinfo{year}{2023}\natexlab{}.
\newblock \bibinfo{title}{MosaicML}.
\newblock \bibinfo{howpublished}{\url{https://www.mosaicml.com}}.   (\bibinfo{year}{2023}).
\newblock


\bibitem[\protect\citeauthoryear{Yang, Wu, Luo, Chiang, Bhardwaj, Kwon, Zhuang, Luan, Mittal, Shenker, et~al\mbox{.}}{Yang et~al\mbox{.}}{2023}]%
        {yang2023skypilot}
\bibfield{author}{\bibinfo{person}{Zongheng Yang}, \bibinfo{person}{Zhanghao Wu}, \bibinfo{person}{Michael Luo}, \bibinfo{person}{Wei-Lin Chiang}, \bibinfo{person}{Romil Bhardwaj}, \bibinfo{person}{Woosuk Kwon}, \bibinfo{person}{Siyuan Zhuang}, \bibinfo{person}{Frank~Sifei Luan}, \bibinfo{person}{Gautam Mittal}, \bibinfo{person}{Scott Shenker}, {et~al\mbox{.}}} \bibinfo{year}{2023}\natexlab{}.
\newblock \showarticletitle{$\{$SkyPilot$\}$: An Intercloud Broker for Sky Computing}. In \bibinfo{booktitle}{{\em 20th USENIX Symposium on Networked Systems Design and Implementation (NSDI 23)}}. \bibinfo{pages}{437--455}.
\newblock


\bibitem[\protect\citeauthoryear{Zaharia, Chen, Davidson, Ghodsi, Hong, Konwinski, Murching, Nykodym, Ogilvie, Parkhe, et~al\mbox{.}}{Zaharia et~al\mbox{.}}{2018}]%
        {zaharia2018accelerating}
\bibfield{author}{\bibinfo{person}{Matei Zaharia}, \bibinfo{person}{Andrew Chen}, \bibinfo{person}{Aaron Davidson}, \bibinfo{person}{Ali Ghodsi}, \bibinfo{person}{Sue~Ann Hong}, \bibinfo{person}{Andy Konwinski}, \bibinfo{person}{Siddharth Murching}, \bibinfo{person}{Tomas Nykodym}, \bibinfo{person}{Paul Ogilvie}, \bibinfo{person}{Mani Parkhe}, {et~al\mbox{.}}} \bibinfo{year}{2018}\natexlab{}.
\newblock \showarticletitle{Accelerating the machine learning lifecycle with MLflow.}
\newblock \bibinfo{journal}{{\em IEEE Data Eng. Bull.\/}} \bibinfo{volume}{41}, \bibinfo{number}{4} (\bibinfo{year}{2018}), \bibinfo{pages}{39--45}.
\newblock


\end{thebibliography}

\end {document}